\documentclass[twocolumn,noshowpacs,preprintnumbers,amsmath,amssymb]{revtex4}

\usepackage{verbatim}
\usepackage{wasysym}

\usepackage{graphicx}
\usepackage{dcolumn}
\usepackage{bm}
\usepackage{wrapfig}

\begin{document}

\title{ZT enhancement in solution-grown Sb$_{(2-x)}$Bi$_{x}$Te$_{3}$ nanoplatelets}

\author{Marcus Scheele}
\email{scheele@chemie.uni-hamburg.de}
\author{Anna-Marlena Kreuziger}
\author{Andreas Kornowski}
\author{Christian Klinke}
\author{Horst Weller}
\affiliation{Institute of Physical Chemistry, University of Hamburg, 20146 Hamburg, Germany}

\author{Niels Oeschler}
\author{Igor Veremchuk}
\affiliation{Max Planck Institute of Chemical Physics of Solids, 01187 Dresden, Germany}

\author{Klaus-Georg Reinsberg}
\author{Jose Broekaert}
\affiliation{University of Hamburg, Institute of Inorganic and Applied Chemistry, 20146 Hamburg, Germany}

\begin{abstract} 

We report a solution-processed, ligand supported synthesis of 15-20 nm thick Sb$_{(2-x)}$Bi$_{x}$Te$_{3}$ nanoplatelets. After complete ligand removal by a facile NH$_{3}$-based etching procedure, the platelets are spark plasma sintered to a p-type nanostructured bulk material with preserved crystal grain sizes. Due to this nanostructure, the total thermal conductivity is reduced by 60 \% in combination with a reduction in electric conductivity of as low as 20 \% as compared to the bulk material demonstrating the feasibility of the phonon-glass electron-crystal concept. An enhancement in the dimensionless thermoelectric figure of merit of up to 15 \% over state-of-the-art bulk materials is achieved meanwhile shifting the maximum to significantly higher temperatures.

\end{abstract}

\maketitle

Recently, Bi$_{2}$Te$_{3}$ based nanostructured materials have received great attention due to their outstanding thermoelectric properties. From the first reports in the 1950s until today, the dimensionless thermoelectric figure of merit (ZT) of such materials at room temperature has been improved threefold. From 0.5 for pure Bi$_{2}$Te$_{3}$ bulk samples \cite{1} over 1.14 for bulk (Bi$_{2}$Te$_{3}$)$_{0.25}$(Sb$_{2}$Te$_{3}$)$_{0.72}$(Sb$_{2}$Se$_{3}$)$_{0.03}$ \cite{2} and 1.2 for nanostructured BiSbTe alloys \cite{3} to 1.56 for nanostructured Sb$_{1.52}$Bi$_{0.48}$Te$_{3}$ \cite{4} with "coherent interfaces", advances in semiconductor manipulation have yielded impressive results in this ecologically highly promising field. ZT is estimated to require a value of 3 to be competitive with conventional cooling devices and to open up novel pathways for efficient and greener power generation.

The record high efficiency of 2.4 was reported for molecular beam epitaxy engineered thin films of Bi$_{2}$Te$_{3}$/Sb$_{2}$Te$_{3}$ layers, which may be difficult to use in large-scale applications but convincingly demonstrated the potential for further improvements to come from nanostructured Bi$_{2}$Te$_{3}$ based materials \cite{5}.

In order to fabricate such materials on a macroscopic scale, one conventionally applies high pressure and suitable temperatures to sinter a Bi$_{2}$Te$_{3}$ based nanopowder to a dense nanocomposite with preserved crystal grain boundaries. Such nanocomposites have been studied by the means of transmission electron microscopy (TEM), energy dispersive X-ray spectroscopy (EDS) \cite{6} and scanning electron microscopy (SEM) \cite{7}. It is believed that the unique structural details in these materials such as laminated structure, coherent interfaces, nanoprecipitates with defect concentrations and broad size distribution of crystalline domains effect all three parameters of ZT, namely the thermopower, electric and thermal conductivity and can lead to an overall improvement of thermoelectric efficiency.

Synthetic strategies to Bi$_{2}$Te$_{3}$ based nanopowders can be divided into two approaches: (a) ligandless or (b) ligand supported nanograin growth. Advantages of the former are the absence of organic impurities and the good alloying possibilities by standard semiconductor manipulations. As a matter of this, the ligandless approach was more successful recently and all of the mile stone achievements in enhancing zT as cited above were due this strategy. An instructive summary has been provided recently by Ren and coworkers \cite{8}. However, it is found almost impossible to achieve a good size control and narrow size distribution of nanoparticles by this strategy. This is the major advantage of solution processed, ligand supported nanoparticle growth strategies. Significantly better size control as compared to ligandless approaches has been achieved for a variety of high performance thermoelectric materials including  Bi$_{2}$Te$_{3}$ \cite{7} and Sb$_{(2-x)}$Bi$_{x}$Te$_{3}$ \cite{9}. Majumdar and co-workers have shown that self-assembled colloidal PbSe nanoparticles synthesized in solution show an enhanced thermopower due to sharp spikes in the density of states because of quantum confinement effects \cite{10}. We have reported recently a procedure capable of completely removing the ligand sphere of formally organically protected nanocrystals \cite{11}. Thus treated nanoparticles sintered to a macroscopic composite show the same electric conductivity as the bulk material. This tool in combination with the advances in solution processed nanotechnology opens up pathways to thermoelectric studies of nanocomposites with grain sizes where low-dimensional effects are really prominent, that is drastically below 100 nm. Only when size and size-distribution of nanoparticles are small, one can obtain detailed insight into the change in thermoelectric parameters due to the limited dimensions. In the recent past, a growing number of theoreticians have turned to modelling the thermoelectric properties of such nanogranular composites, including the thermoelectric coefficient and figure of merit \cite{12}, the transport properties \cite{13} and the power factor \cite{14}. Their results suggest that nanogranular materials not only display the easily understood decrease in lattice thermal conductivity but also alterations in transport properties which are not as straight forward to comprehend as the influence on phonon transport. For the on-going discussion to advance, new systems need to be designed to deliver experimental data for comparison with theoretical predictions. In this respect, ligand supported growth of nanostructures provides an additional tool to access new and complex thermoelectric materials  which are difficult to obtain by conventional top-down approaches.

In this letter, we report the large scale synthesis of Sb$_{(2-x)}$Bi$_{x}$Te$_{3}$ nanoplatelets in solution, their purification from organic ligands, the fabrication of macroscopic nanocomposites and their full thermoelectric characterization. 

Sb$_{(2-x)}$Bi$_{x}$Te$_{3}$ nanoplatelets are synthesized similarly to a previously developed protocol for Bi$_{2}$Te$_{3}$ nanoparticles \cite{11}. The acetates of Bi$^{3+}$ and Sb$^{3+}$ are treated with excess 1-dodecanethiol (DDT) under vacuum to remove acetic acid and form the metal thiolates which are easily soluble in organic media. Injecting the mild reducing agent oleylamine into this solution exclusively initiates the formation of a Bi$^{0}$ species, referred to in the following as "slow reduction". Adding a solution of tellurium in trioctylphosphine (Te@TOP) within minutes after initiating the slow reduction triggers a much faster reduction by the more potent reducing agent trioctylphosphine (TOP). The Me$^{0}$ species formed during the "fast reduction" is unstable in the presence of the tellurium complex and reacts to the ternary antimony-bismuth-telluride compound. Shortly after beginning the fast reduction under these conditions, thin nanostructures only few nanometers in thickness of Sb-Bi-Te can be found, referred to in the following as "nanoflakes". If kept at moderate temperatures (60$^{\circ}$C), the nanoflakes will combine to larger structures, referred to in the following as "nanosheets". During this process the thickness of the sheets does not change significantly. Over time, an increasing number of defined geometric features like sharpe edges, corners and even symmetric hexagons is observed. Typically, the dimensions of the final nanosheets are 50-200 nm across and up to 5 nm in thickness as estimated by X-ray powder diffraction (XRPD) measurements. By increasing the reaction temperature (90$^{\circ}$C vs. 60$^{\circ}$C), the thin and often porous nanosheets grow in thickness by the up-take of Sb$_{2}$Te$_{3}$ from the solution. The nanostructures resulting from this process are single-crystalline, 15 - 20 nm in thickness and will be referred to in the following as "nanoplatelets". 

In Fig. 1, transmission electron microscopy (TEM) reveals the shape of typical nanosheets (1a) and nanoplatelets (1b). A high resolution (HR-TEM) image (1c) verifies the discontinuous nature of the nanosheets with the amorphous carbon substrate visible underneath. It is noteworthy that the whole sheet appears to be almost single-crystalline which is further studied in Fig. 1e. Here, the fast Fourier transformed (FFT) of the image displays a clear atomic ordering, however with a slight directional misalignment of the individual crystalline domains. From this, it is believed that the porous nanosheets evolve from nanoflakes throughout the course of the reaction. During this process, the individual nanoflakes have to align perfectly in one crystalline direction to eventually form a single crystal. The crystal depicted in the inset of FIg. 1c is therefore an intermediate stage in this process. In contrast, the thicker nanoplatelets are continuous and perfectly single-crystalline (see Fig. 1d). 

\begin{figure}[htbp]
  \centering
  \includegraphics[width=0.45\textwidth]{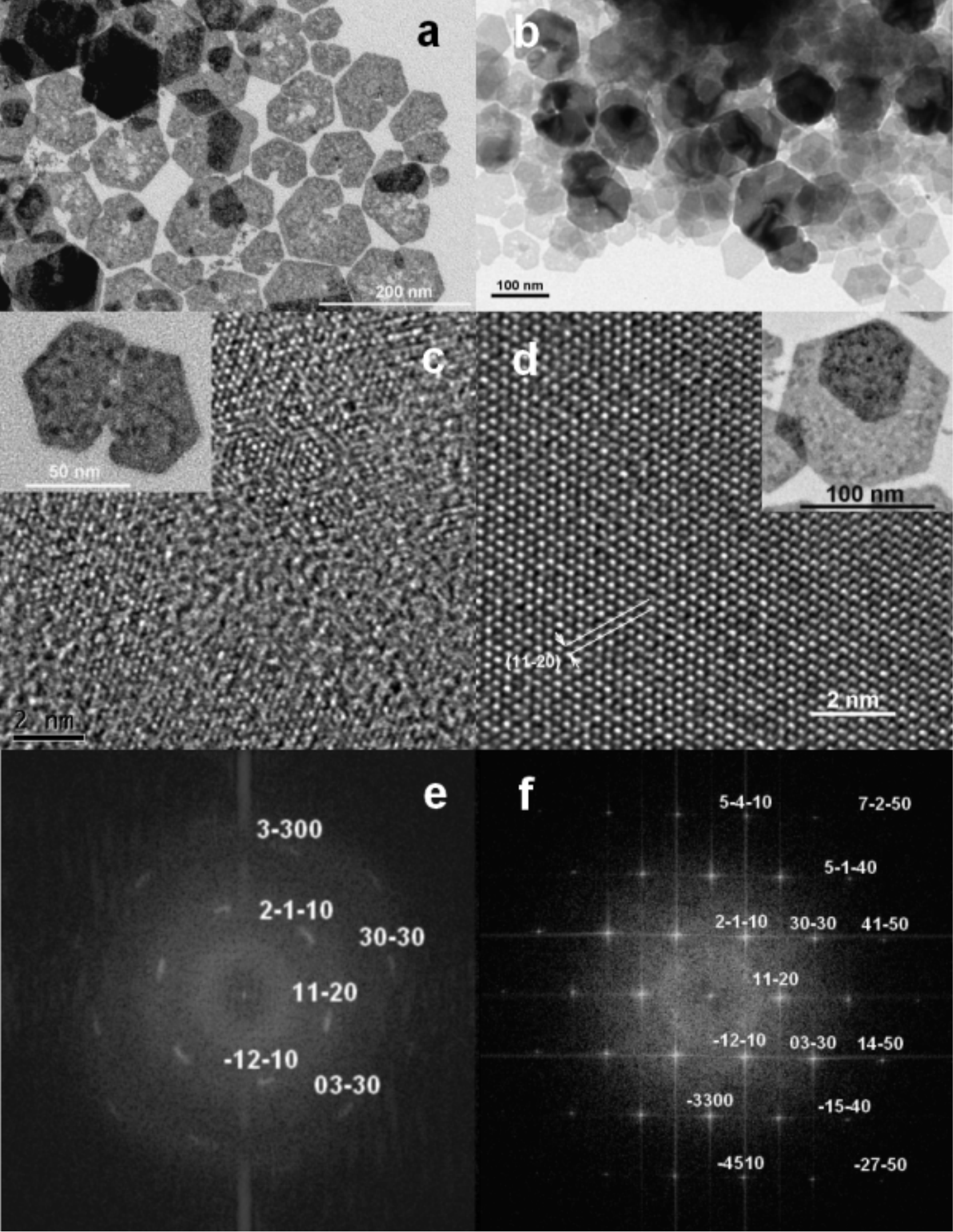}
  \caption{\textit{TEM images of $Sb_{(2-x)}Bi_{x}Te_{3}$ nanosheets (a) and nanoplatelets (b). HR-TEM images of individual (see inset) nanosheets (c) and nanoplatelets (d). The indicated lattice spacing is 2.15 $\AA$ which corresponds to the spacing of (11-20) planes. Fast Fourier transformed of the HR-TEM images of nanosheets (e) and nanoplatelets (f). In each case, the direction of view is $\left[ 0001 \right]$.}}
\end{figure}

As visible in the FFT of the HR-TEM image (1f), the direction of growth of the nanoplatelets is in the a-b-plane, that is along the $\left[ 110 \right]$ direction ($\left[ 11-20 \right]$ in Bravais annotation). This is a typical behaviour of Bi$_{2}$Te$_{3}$ and Sb$_{2}$Te$_{3}$ owed to its highly anisotropic crystal structure \cite{15,16}.

Compositional analysis is shown in Fig. 2 by the means of XRPD, energy dispersive X-ray spectroscopy (EDS) and inductively plasma coupled optical emission spectroscopy (ICP-OES). The XRPD (Fig. 2 left) of Sb$_{(2-x)}$Bi$_{x}$Te$_{3}$ nanoplatelets reveals a single-phase product with slightly broadened reflections typical for crystals with nanoscalic dimensions. Position and intensity of the reflections can be attributed to an intermediate of Bi$_{2}$Te$_{3}$ and Sb$_{2}$Te$_{3}$ with a Sb:Bi ratio much larger than unity. Especially important in this respect are the (015) (here: 2 $\theta$ = 28.2$^{\circ}$) and the (110) (here 2 $\theta$ = 42.3$^{\circ}$) reflections which compares to 27.6$^{\circ}$ vs. 28.4$^{\circ}$ and 42.0$^{\circ}$ vs. 42.4$^{\circ}$ for Bi$_{2}$Te$_{3}$ and Sb$_{2}$Te$_{3}$, respectively. Note: Since bismuth has a larger Bohr radius than antimony, the lattice spacing in Bi$_{2}$Te$_{3}$ is slightly larger than in Sb$_{2}$Te$_{3}$. Hence the deviations in XRPD despite the identical space group R-3m (166).

\begin{figure}[htbp]
  \centering
  \includegraphics[width=0.45\textwidth]{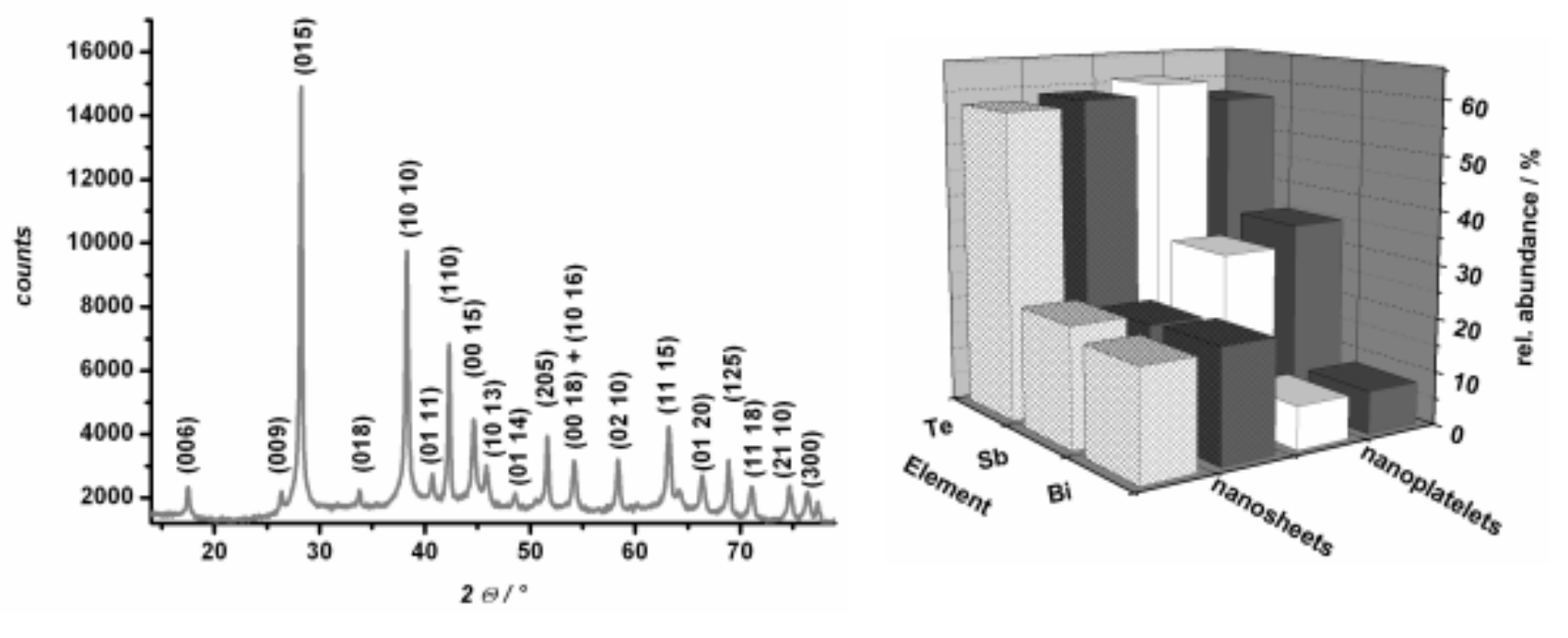}
  \caption{\textit{(Left) XRPD of Sb$_{(2-x)}$Bi$_{x}$Te$_{3}$ nanoplatelets with Sb:Bi $\approx$ 4:1. Indexing according to space group R-3m (166). (Right) Elemental analysis of Sb$_{(2-x)}$Bi$_{x}$Te$_{3}$ nanosheets (hatched) and nanoplatelets (blank). Individual nanocrystals were analyzed by EDS (light), large quantities of nanocrystal powder were analyzed by ICP-OES (dark).}}
\end{figure}

EDS and ICP-OES analysis (Fig. 2 right) reveal antimony, bismuth and tellurium as the main constituting elements in the nanosheets and nanoplatelets. When comparing the quantitative elemental analysis of nanosheets and nanoplatelets it becomes evident that the Sb$_{(2-x)}$Bi$_{x}$Te$_{3}$ sheets have a much larger bismuth content (Sb:Bi $\approx$ 1:1, Sb$_{1.0}$Bi$_{1.0}$Te$_{3.0)}$ than the platelets (Sb:Bi $\approx$ 4:1, Sb$_{1.7}$Bi$_{0.4}$Te$_{3.0}$). Note that optimized, bulk Sb${(2-x)}$Bi$_{x}$Te$_{3}$ usually has a Sb:Bi ratio of 3:1 (Sb$_{1.5}$Bi$_{0.5}$Te$_{3}$) \cite{3}. In both of the two structures the relative abundance of tellurium is always 57-59 at-\%. The analytical results of individual nanosheets and platelets obtained by EDS are in good agreement with the results of ICP-OES analysis of the same material subsequent to wet chemical digestion showing the homogeneity in chemical composition of the individual nanocrystals.

In the next step, all organic residues are removed from the inorganic nanosheets or nanoplatelets to allow for high electric conductivities. Where solvents can be effectively withdrawn by multiple washing steps, the separation from the stabilizing agent DDT is realized by a modified procedure similar to a previously developed protocol \cite{11}. As the only alteration, after ligand exchange with oleic acid we apply a methanolic NH$_{3}$ solution rather than hydrazine hydrate. The advantage is the non-reductive nature of NH$_{3}$ in comparison to the powerful reducing agent hydrazine. This way, an unwanted partial reduction of the nanosheets during the washing procedure can be excluded. After drying the inorganic material under vacuum, we obtain a dark-grey nanopowder. According to XRPD, this procedure has no significant impact on the crystalline phase of the nanomaterial.

To fabricate a nanostructured bulk material, this powder is spark plasma sintered (SPS) to a macroscopic pellet. Pellets of Sb$_{(2-x)}$Bi$_{x}$Te$_{3}$ nanosheets or nanoplatelets are silver-metallic in appearance with a density of 5.65 - 5.85 g cm$^{-3}$ (83 - 86 \% of theoretical density) under the conditions specified in the experimental section.

In the following, we will focus on the thermoelectric properties of Sb$_{1.7}$Bi$_{0.4}$Te$_{3.0}$ nanoplatelets.

In Fig. 3, the fine structure of the sintered pellets and its impact on the thermal conductivity is investigated. As displayed by SEM imaging in Fig. 3a, the shape of individual nanoplatelets is preserved in the final pellets yielding a highly polycrystalline, layered material. We compare our thermal conductivity results with state-of-the-art macrocrystalline, bulk Sb$_{1.5}$Bi$_{0.5}$Te$_{3}$ \cite{3}.

\begin{figure}[htbp]
  \centering
  \includegraphics[width=0.45\textwidth]{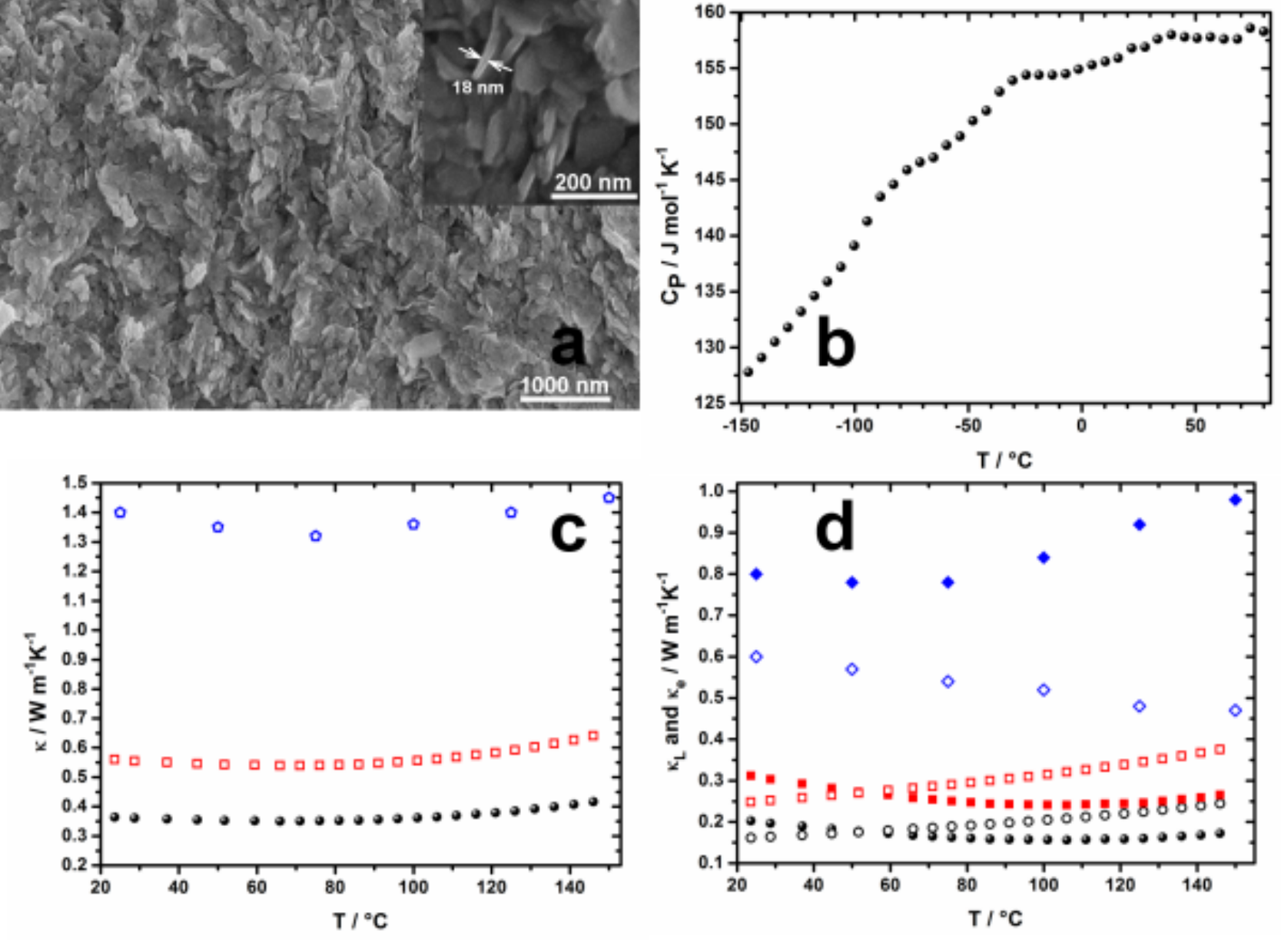}
  \caption{\textit{(a) SEM image of a spark plasma sintered pellet of Sb$_{1.7}$Bi$_{0.4}$Te$_{3.0}$ nanoplatelets. The inset contains a side view on individual nanoplatelets revealing their approximate thickness. (b) Specific heat capacity of the nanoplatelet. (c) Total thermal conductivity of nanoplatelets uncorrected (closed circles), corrected for porosity (open squares) and of a Sb$_{1.5}$Bi$_{0.5}$Te$_{3}$ bulk ingot taken from reference \cite{3} (open diamonds). (d) Lattice ($\kappa_{L}$, closed) and electronic ($\kappa_{e}$, open) thermal conductivity of nanoplatelets uncorrected (circles), corrected for porosity (squares) and bulk ingots (diamonds).}}
\end{figure}

To calculate the total thermal conductivity ($\kappa$) from Laserflash measurements, we measured the specific heat (C$_{P}$, Fig. 3b) repeatedly of several samples and on using three different measurement systems (Physical Property Measurement System by Quantum Design; Differential Scanning Calorimeter by Netzsch and Linseis). The deviation of all measurements was less than 5 \%. At 300 K we obtain C$_{P}$ = 158 J mol$^{-1}$ K$^{-1}$ or 235 J kg$^{-1}$ K$^{-1}$, respectively. With this, an average density ($\rho$) of 5.73 g cm$^{-3}$ and the thermal diffusivity results (D$_{t}$) we calculate the total thermal conductivity following the equation: 

\begin{equation}
	\kappa = D_{t} \: \rho \: C_{P}
\end{equation}

When measuring the transport properties of Bi$_{2}$Te$_{3}$-based compounds, great care must be taken about the individual measurement directions relative to the crystal's orientation. As Fleurial et al. have worked out, the anisotropy in thermal and electric transport between the a-b-plane and the c-axis of single-crystalline Bi$_{2}$Te$_{3}$ is 2 to 2.5 \cite{17}. A similar degree of transport anisotropy can be expected for single-crystalline SbBiTe$_{3}$. In pressed pellets of SbBiTe$_{3}$ nanoparticles, transport anisotropy depends on the degree of ordering of the individual nanocrystals in the pellet. As Ma et al. have shown, the random orientation of nanocrystals during hot-pressing can decrease transport anisotropy in nanostructured bulk SbBiTe$_{3}$ to less than 5 \% \cite{18}. For the material presented in this work, we determined a temperature dependent anisotropy in the thermal conductivity of 8 to 18 \% with the lower thermal conductivity in the c-direction. This anisotropy has been taken into account for all thermal conductivity measurements discussed below.

To allow a quantitative comparison of transport properties with the previous reports mentioned above \cite{3,4}, it is important to account for different degrees of porosity in the measured samples. Where Poudel et al. and Xie et al. measured samples with 100 \% and 96 \% relative density, respectively, the material presented in this work shows relative densities of only 83 to 86 \%. As several groups have reported, porosity decreases thermal conductivity \cite{19} and electric conductivity \cite{20} alike. Where the effect of porosity on ZT is usually small if not negligible \cite{19}, individual transport parameters must be corrected for porosity in order to judge the effect of nanostructuring alone. To do so, we follow Adachi et al. by using a Maxwell-Eucken expression \cite{21}:

\begin{equation}
	x_{P} = x_{0} \cdot \frac{1 - P}{1 + \beta P}
\end{equation}

where $x$ is either the thermal or the electric conductivity, $x_{P}$ is the transport property in the porous medium, $x_{0}$ is the same property in the 100 \% dense medium, $P$ is the degree of porosity (a fraction between 0 and 1) and $\beta$ an empirical parameter describing the shape of the pores. In the following, we set $P$ to 0.15 and $\beta$ to 2.0 which is a fair estimate related to similar works reported in literature \cite{20}. Thus, the porosity in the material presented is estimated to account for a reduction in both, thermal and electric conductivity, to 65 \% of the theoretical value to be expected for an absolutely dense sample. Comparisons with other models accounting for the effects of porosity (eg. the percolation model \cite{22}) verify this assumption to be very reasonable \cite{23}. To display the pure effect of nanostructuring on the transport properties, we will use only the porosity corrected values in the following discussion. In Fig. 3 and 4 we prefer to show both, the uncorrected and corrected measurements, for clarity.

The qualitative behaviour of $\kappa(T)$ of the pellets of nanoplatelets (Fig. 3c) is similar to the bulk material in so far that the variation over the measurement range from 25$^{\circ}$C to 145$^{\circ}$C is less than 10 \% with a minimum at 80$^{\circ}$C. Quantitatively however, with 0.56 W m$^{-1}$K$^{-1}$ at 300 K, $\kappa$ is 60 \% lower than that of a comparable bulk material \cite{3}. Further, it is 50 \% lower than nanostructured bulk Sb$_{1.5}$Bi$_{0.5}$Te$_{3}$ obtained by ball-milling \cite{3} and 15 \% lower than a melt-spun nanostructured bulk sample \cite{4}. As discussed below, some of this decrease results from a simultaneous decrease in electric conductivity ($\sigma$) (see Fig. 4) which overall has no effect on the thermoelectric efficiency due to the Wiedemann-Franz law. Therefore, a more suitable measure to judge the potential of our material for improved thermoelectrics is the lattice part ($\kappa_{L}$) of the thermal conductivity. To calculate $\kappa_{L}$ at a given temperature (T), we follow Tritt and coworkers and use L = 2.0 $\cdot$ 10$^{-8}$ V$^{2}$ K$^{-2}$ which is accepted to be the appropriate value of the Lorentz number for heavily degenerated semiconductors. 

\begin{equation}
	\kappa_{L} = \kappa - \kappa_{e} = \kappa - L  \: \sigma \: T
\end{equation}

With $\kappa_{L}$ = 0.31 W m$^{-1}$ K$^{-1}$ at 300 K (Fig. 3d), the material in this communication shows a decrease in phononic heat transport by 60 \% with reference to the bulk material, 10 \% with reference to nanostructured bulk samples by ball-milling and an increase by 20 \% compared to nanostructured melt-spun samples. Another important aspect in Fig. 3d is the different behaviour of  $\kappa_{L}(T)$ and  $\kappa_{e}(T)$ in the nanostructured material as compared to the bulk state. At room temperature, heat transport in the nanoplatelets is dominated by lattice vibrations, that is, $\kappa_{L}$ $>$  $\kappa_{e}$. With increasing T, $\kappa_{e}$ increases and $\kappa_{L}$ decreases so that at 145$^{\circ}$C, $\kappa_{e}$ is one and a half times larger than $\kappa_{L}$. Quite the opposite behaviour is known about bulk Sb$_{1.5}$Bi$_{0.5}$Te$_{3}$ where $\kappa_{e}$ is always smaller than $\kappa_{L}$ in this temperature regime. 

In Fig. 4, we display our $\sigma$ (4a) and thermopower ($S$, 4b) measurements, combine it to the power factor ($\sigma S^{2}$, 4c) and calculate the dimensionless figure of merit ($ZT$, 4d) according to

\begin{equation}
 ZT = \frac{\sigma S^{2}}{\kappa} T	
\end{equation}

\begin{figure}[htbp]
  \centering
  \includegraphics[width=0.45\textwidth]{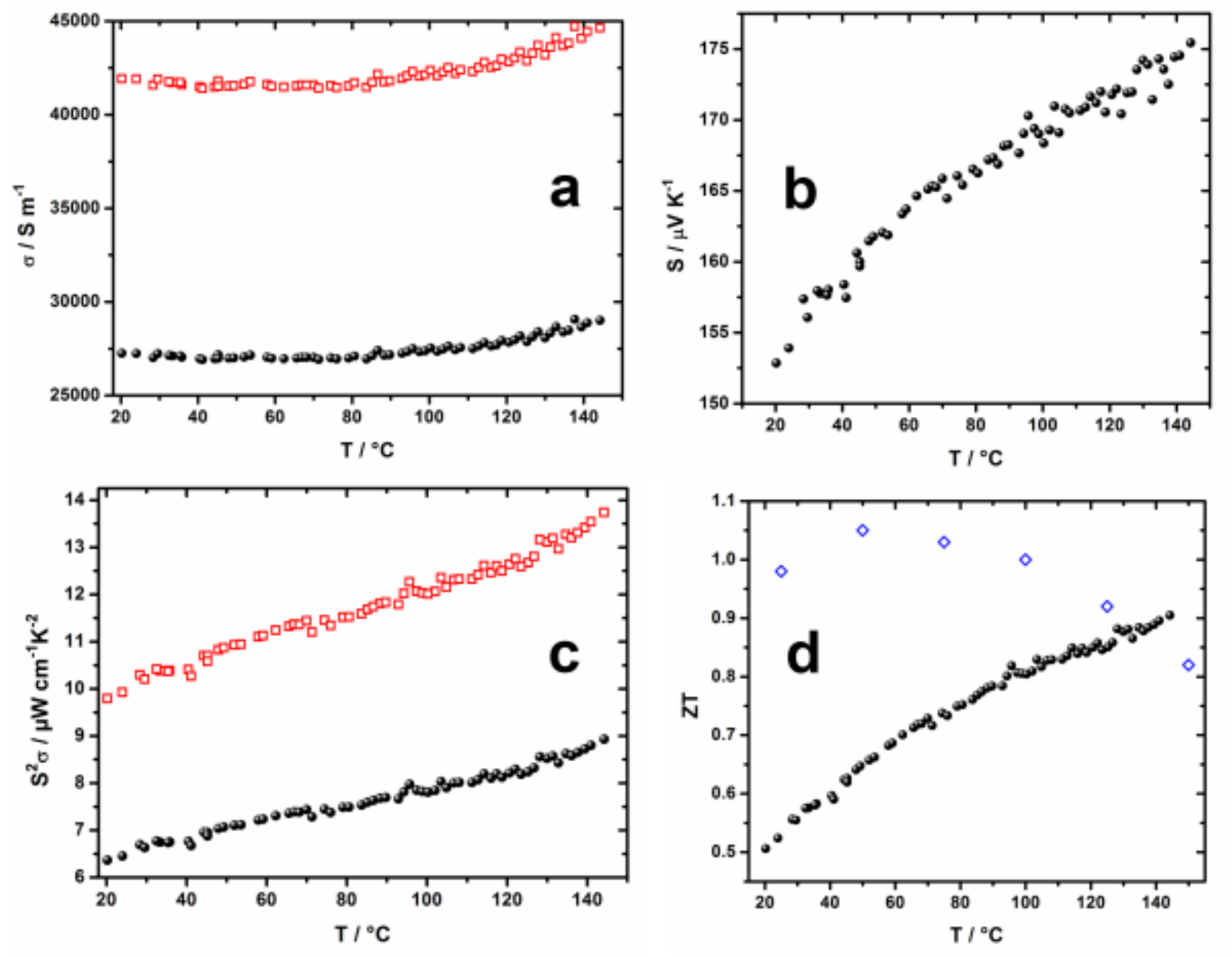}
  \caption{\textit{(a) Electric conductivity, (b) thermopower, (c) power factor and (d) dimensionless figure of merit of a pellet of Sb$_{1.7}$Bi$_{0.4}$Te$_{3}$ nanoplatelets uncorrected (closed circles) and corrected for porosity (open squares). For comparison, the figure of merit of bulk Sb$_{1.5}$Bi$_{0.5}$Te$_{3}$ ingots \cite{4} is displayed (open diamonds).}}
\end{figure}

At room temperature, $\sigma$ is decreased to 40 \% of the bulk value which is the result of electron scattering at crystal grain boundaries. The large surface-to-volume ratio of nanograins results in high trap-state densities at each grain boundary caused by defects and dangling bonds. These trap-states immobilize charge carriers, leading to a reduced, effective mobility as compared to single crystals \cite{24}. Further, the grain boundaries become electrically charged and provide a grain boundary potential barrier to electric transport \cite{25}. In contrast to bulk $Sb_{1.5}Bi_{0.5}Te_{3}$, the pellets of nanoplatelets show a positive $\Delta \sigma$/$\Delta T$ ratio. This behaviour has recently been predicted by Nolas and co-workers \cite{26} for nanostructured materials where transport properties are dominated by grain boundary potential scattering in combination with phonon scattering. At not too low temperatures, electric transport across the grain boundaries can be described by thermionic emission with $\sigma (T)$ $\sim$ $T^{-1/2} \exp(-E_{B} / kT)$ where $E_{B}$ is the height of the grain boundary potential barrier. Thus, for certain temperatures and grain boundary potential barriers, $\Delta \sigma$/$\Delta T$ can become positive. This is an important difference to bulk , the pellets of nanoplatelets show a positive $\Delta \sigma$/$\Delta T$ ratio. This behaviour has recently been predicted by Nolas and co-workers \cite{26} for  where the negative $\Delta \sigma$/$\Delta T$ ratio forces a maximum in ZT at 50$^{\circ}$C preventing an efficient application at significantly higher temperatures.

The positive thermopower (Fig. 4b) indicates p-type behaviour in accordance with other Sb$_{1.5}$Bi$_{0.5}$Te$_{3}$ materials. The magnitude of, at best, 175 $\mu$V K$^{-1}$ is lower than that of comparable materials by other groups which can reach up to 220 $\mu$V K$^{-1}$. Like no other thermoelectric parameter, $S$ varies strongly with the chemical composition. For example, for the nanosheets displayed in Fig. 1a,c,e with a composition of Sb$_{1.0}$Bi$_{1.0}$Te$_{3}$ (see Fig. 2b), we find a thermopower of merely 30 $\mu$V K$^{-1}$. We note that the composition of the nanoplatelets (with S = 175 $\mu$V K$^{-1}$) is with Sb$_{1.7}$Bi$_{0.4}$Te$_{3}$ too rich in antimony to fully compete with the optimized Sb$_{1.5}$Bi$_{0.5}$Te$_{3}$ samples applied in other works showing the best thermopower values. We believe that further optimization of our synthesis towards this ideal chemical composition will also lead to thermopower values of up to 220 $\mu$V K$^{-1}$. The precondition that the thermopower of nanostructured materials can be at least as high as in single crystals has been verified by several other groups already \cite{27,28,29,30}.

The combination of the two parameters to the power factor ($\sigma S^{2}$, 4c) leads to an unusual temperature dependence. In bulk samples, $\sigma S^{2}$ decreases with T at not too low temperatures owing to the negative $\Delta \sigma$/$\Delta T$ ratio. This also applies to the nanostructured samples reported in \cite{3,4}. In contrast, for the material in this work, $\sigma S^{2}(T)$ increases with temperature since  $\Delta \sigma$/$\Delta T$ is positive.

Consequently, $ZT(T)$ (Fig. 4d) also increases with temperature. Starting with a moderate $ZT_{25^{\circ} C} = 0.5$, at the highest measurement temperature we find $ZT_{145^{\circ} C} = 0.9$ which is 15 \% larger than that of a comparable bulk material. A maximum in $ZT$ can be estimated to be found significantly above 145$^{\circ}$C when thermopower begins to be depleted by bipolar conduction and other effects. This is in agreement with other reports \cite{3,18,27} which translated the maximum in $ZT$ of bulk $SbBiTe_{3}$ at 50$^{\circ}$C to about 100$^{\circ}$C via nanostructuring. The present work demonstrates the continuation of this trend probably due to the introduction of larger grain boundary potential barriers which has been shown to drastically alter electron transport. This needs to be verified in the future via Hall mobility measurements.  

In Fig. 5(left), the reason for the enhanced thermoelectric efficiency is summarized. By fitting the experimental data on the transport properties of bulk and nanostructured $Sb_{(2-x)}Bi_{x}Te_{3}$ from Fig. 3c, d and 4a, we plot the relative reduction of $\sigma$, $\kappa$ and $\kappa_{L}$ as a function of $T$ resulting from the material's nanostructure. Over the entire temperature range, the reduction in $\kappa$ is larger than the reduction in $\sigma$. As mentioned above, this is the key requirement for designing more efficient thermoelectric materials. 

It is noteworthy that the relative reduction of $\kappa$ is with 60 \% practically constant over the entire temperature regime. This is the result of the increased relative electric conductivity in combination with a simultaneous decrease in the relative lattice thermal conductivity. Materials with this behaviour are often referred to as "phonon-glass electron-crystals" (PGEC), meaning materials with good charge carrier but poor phonon transport.

\begin{figure}[htbp]
  \centering
  \includegraphics[width=0.45\textwidth]{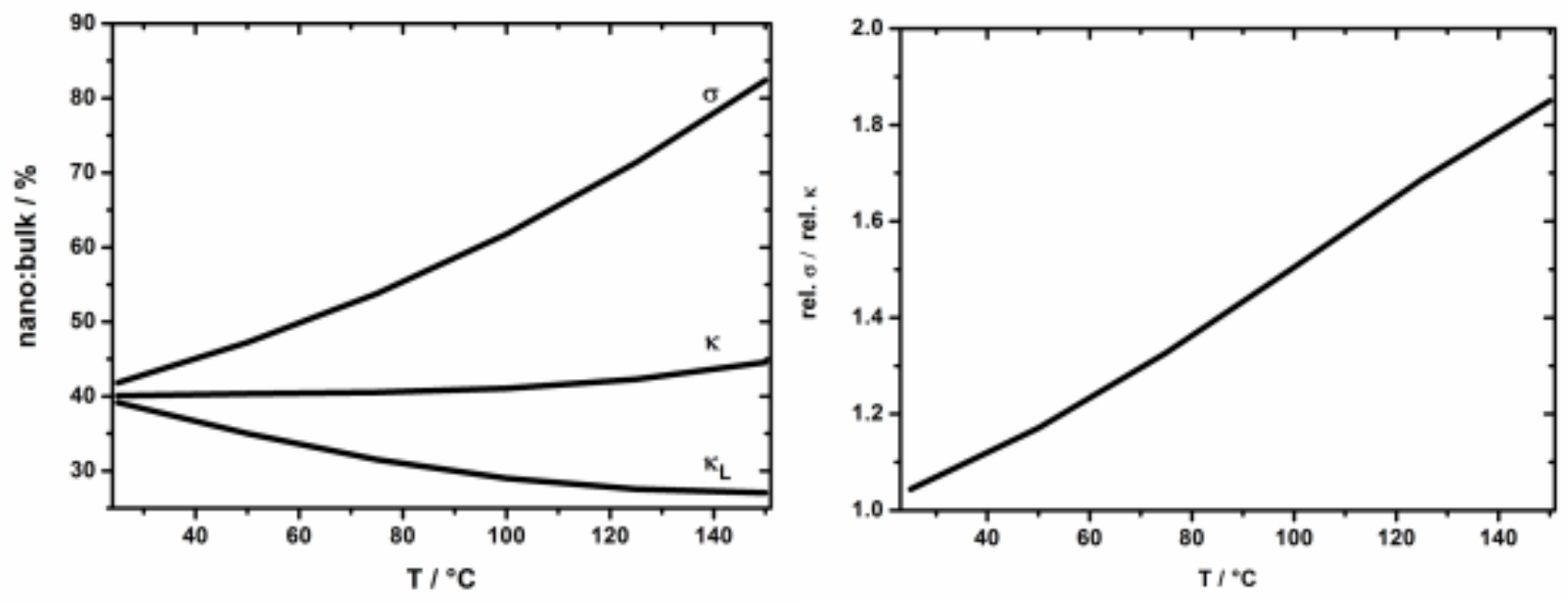}
  \caption{\textit{(Left) Relative decrease of transport parameters of a pellet of $Sb_{1.7}Bi_{0.4}Te_{3.0}$ nanoplatelets as compared to bulk $Sb_{1.5}Bi_{0.5}Te_{3.0}$ ingots. (Right) "PGEC-factor" (the ratio of relative electric to total thermal conductivity) of $Sb_{1.7}Bi_{0.4}Te_{3.0}$ nanoplatelets.}}
\end{figure}

In Fig. 5 (right) we introduce the temperature dependent "PGEC-factor" defined as the ratio of the relative electric to total thermal conductivity compared to the bulk state. For the pellets of $Sb_{1.7}Bi_{0.4}Te_{3.0}$ nanoplatelets in this communication, the PGEC-factor starts at almost unity at room temperature but increases quickly to 1.8 at 140$^{\circ}$C. Provided the thermopower can be improved to the bulk value by adjusting the chemical composition to $Sb_{1.5}Bi_{0.5}Te_{3.0}$, the PGEC-factor is a direct measure ($ZT_{nano}$ = $ZT_{bulk}$ x PGEC-factor) for the expected improvement in $ZT$ as compared to the bulk material. This way, $ZT$ = 1.5 is within reach.

In conclusion, we have demonstrated the applicability of ligand supported, bottom-up synthesized $Sb_{(2-x)}Bi_{x}Te_{3}$ nanoplatelets for highly efficient, macroscopic, p-type thermoelectric materials. The enhancement in $ZT$ is achieved by the phonon-glass electron-crystal effect which describes the preferential scattering of phonons as compared to electrons by the nanostructure. The key to this property is a reduction in the lattice (or phononic) thermal conductivity leading to an enhancement in $ZT$ by up to 15 \% as compared to comparable bulk materials. Future adjustments of the chemical composition hold for an enhancement of up to 80 \%. The maximum in ZT is shifted to larger temperatures which seems to be a direct consequence of the nanostructure. 

\

\textbf{ACKNOWLEDGEMENTS}

\

We thank Katrin Meier for help with SPS experiments, William Töllner for help with DSC measurements and Alexander Littig for fruitful discussions. A PhD-grant by the Studienstiftung des deutschen Volkes is gratefully acknowledged.  

\

\textbf{EXPERIMENTAL DETAILS}

\

All manipulations were carried out under an inert atmosphere using standard Schlenck techniques if not stated otherwise. 

\

\textit{(I) Preparation of a 0.500 M solution of tellurium in TOP (Te@TOP)}

\

In a glovebox, tellurium (1.276 g, 10.00 mmol, 99.999 \%, Chempur) and tetradecylphosphonic acid (102 mg, Alfa Aesar) were suspended in distilled TOP (20.0 mL, 90 \%, Merck) under stirring. It was heated stepwise to 230$^{\circ}$C from room temperature by increasing the temperature by approximately 50$^{\circ}$C every thirty minutes. The final temperature was kept until a completely transparent, orange solution was obtained which turned to bright-yellow on cooling to room temperature. The solution was stored in the glovebox.

\

\textit{(II) Synthesis of Sb1.0Bi1.0Te3.0 nanosheets}

\

In a typical synthesis, bismuth acetate (0.045 g, 0.12 mmol, 99 \% Aldrich) and antimony acetate (0.323 g, 1.08 mmol, 99 \% Aldrich) were mixed with 1-dodecanethiol (13.3 mL, 98 \% Aldrich) and heated to 45$^{\circ}$C for 45 min under vacuum on which a transparent, yellow solution was obtained. The flask was flooded with nitrogen, set to ambient pressure and it was heated to 60$^{\circ}$C on which oleylamine (26.7 mL, 70 \%, Aldrich) was quickly added under stirring (referred to as "slow reduction" in the main body of the paper). After three minutes when the solution had visibly darkened, 3.6 mL of (I) were injected under vigorous stiring (referred to as "fast reduction" in the main body of the paper). After 24 h the as prepared $Sb_{1.0}Bi_{1.0}Te_{3.0}$ nanosheets were ready for further manipulations.

\

\textit{(III) Synthesis of Sb1.7Bi0.4Te3.0 nanoplatelets}

\

The amounts and procedure were identical to (II) with the only alteration being the reaction temperature which was 90$^{\circ}$C instead of 60$^{\circ}$C.

\

\textit{(IV) Purification of Sb(2-x)BixTe3 nanosheets or nanoplatelets for characterization }

\

A fraction of the dark-grey suspension obtained under (II) or (III) was mixed with ethanol (25 vol-\%, analytical grade, Fluka) and centrifuged at 4500 rpm for 5 min. The light yellow supernatant was removed under nitrogen and the almost black precipitate suspended in a few drops of chloroform (analytical grade, Fluka) on which the washing cycle was repeated two more times. The purified nanosheets or nanoplatelets should be stored in the absence of oxygen to prevent aging. 

\

\textit{(V) Ligand removal from Sb(2-x)BixTe3 nanosheets or nanoplatelets}

\

The purified (II) or (III) were precipitated again with ethanol and the supernatant removed after centrifugation. The black precipitate was mixed with a large excess of oleic acid ($\sim$3 mL, 90 \%, Aldrich) and allowed to stir overnight on which a black suspension was formed. The supernatant was removed after short centrifugation and fresh oleic acid was added on which the mixture was allowed to stir for several hours. The supernatant was removed again after centrifugation and it was washed three times with hexane (analytical grade, Aldrich). The precipitate was suspended in a solution of $NH_{3}$ in methanol (2 mL, 7 N, Aldrich). After stirring overnight, the supernatant was removed after centrifugation (4500 rpm, 5 min) and it was washed two times with fresh NH3 in methanol followed by three washing steps with hexane. 

All solvents were removed and it was dried under vacuum overnight on which a fine black powder was obtained.

Typically, the starting amounts specified under (II) yield approximately 100 mg of $Sb_{1.0}Bi_{1.0}Te_{3.0}$ nanosheets and 180 mg of $Sb_{1.7}Bi_{0.4}Te_{3.0}$ nanoplatelets for (III), respectively.

\

\textit{(VI) Fabrication of Sb(2-x)BixTe3 nanosheet pellets by spark plasma sintering}

\

Typically, 100 mg of (V) kept under argon were loaded into a WC/Co die of 8.0 mm x 1.5 mm in area. The powder was pressed to a solid pellet of equal dimensions and approximately 1.5 mm in height by spark plasma sintering in a SPS-515 ET/M apparatus (Dr. Sinter(R) lab). On applying 340 MPa (for rectangular bars) or 530 MPa (for disks) pressure, the die containing the nanopowder was heated from 20$^{\circ}$C to 50$^{\circ}$C in 5.0 min with 10.0 min hold time by applying a DC current between 0 - 165 A and immediately allowed to cool down to room temperature. The obtained $Sb_{(2-x)}Bi_{x}Te_{3}$ nanosheet pellets were mechanically robust and silver-metallic in appearance. 

For thermal conductivity measurements, 170 mg of (V) were loaded into a disk-shape die of 6 mm in diameter to obtain a tablet of $Sb_{(2-x)}Bi_{x}Te_{3}$ nanosheets with 1.3 mm in height. 

\

\textit{Characterization:}

\

(HR-)TEM imaging was performed with a JEOL JEM 2200 FS (UHR) with CESCOR and CETCOR corrector at an acceleration voltage of 200 kV or a JEM-Jeol-1011 microscope at 100 kV with a CCD camera. SEM images were obtained on a LEO1550 scanning electron microscope with a spatial resolution of $\sim$1 nm. XRPDs were recorded using a Philipps X`Pert-diffractometer with Bragg-Brentano-geometry on applying copper-K$_{\alpha}$ radiation ($\lambda$ = 154.178 pm, U = 45 kV; I = 40 mA).

For measurements of the thermopower and resistivity a ZEM-3 apparatus (ULVAC-RIKO) was applied under a low-pressure helium atmosphere. The thermopower was determined by a static dc method where the resistivity was simultaneously measured by a four-terminal set-up.

The specific heat was measured by a relaxation technique in a Physical Property Measurement System by Quantum Design. A heat pulse of 2 \% of the bath temperature has been applied and repeated 3 times at each temperature.

Thermal diffusivity measurements were recorded with a Netzsch LFA-441 and a Netzsch LFA-457 Microflash with a Pyroceram standard for calibration. 

ICP-OES analysis was performed with a Spectro Ciros CCD (Spectro Analytical Instruments) subsequent to powder sample digestion in a microwave assisted sample decomposition system (MARS 5, CEM Corporation) with a mixture of 20 \% nitric acid and tartaric acid (L(+)-tartaric acid, p.a., 99.5 \%, Sigma-Aldrich). The latter proved necessary as complexing agent to overcome the formation of insoluble $Sb_{2}O_{3}$. The Bismuth, antimony and tellurium contents were determined by calibration with matrix matched solutions produced form ICP-standard solutions (1000 mg L$^{-1}$ $Bi$, $Bi(NO_{3})_{3}$ in $HNO_{3}$ 2-3 \%, Merck, 1000 mg L$^{-1}$ $Sb$, $Sb_{2}O_{3}$ in $HCl$ 7 \%, Merck , 1000 mg L$^{-1}$ $Te$, $H_{6}TeO_{6}$ in $HNO_{3}$ 2-3 \%, Merck). The relative errors of the analysis by ICP-OES were $<$ 3.7 \%.

\end{document}